\documentclass{svjour3}                     
\smartqed  
\usepackage{graphicx}
\usepackage{mathptmx}      
\usepackage{bm}
\newcommand{\EQ}{\begin{equation}}
\newcommand{\EN}{\end{equation}}
\newcommand{\EQA}{\begin{eqnarray}}
\newcommand{\ENA}{\end{eqnarray}}
\newcommand{\eq}[1]{(\ref{#1})}

\newcommand{\Eq}[1]{equation~(\ref{#1})}

\newcommand{\Sec}[1]{\S\,\ref{#1}}

\newcommand{\Fig}[1]{Figure~\ref{#1}}

\newcommand{\Tab}[1]{Table~\ref{#1}}

\newcommand{\bra}[1]{\langle #1\rangle}

\newcommand{\meanemf}{\overline{\mbox{\boldmath ${\mathcal E}$}} {}}
\newcommand{\meanEMF}{\overline{\mbox{\boldmath ${\mathcal E}$}} {}}
\newcommand{\meanFF}{\overline{\mbox{\boldmath ${\mathcal F}$}} {}}

\newcommand{\meanB}{\overline{B}}

\newcommand{\meanAA}{\overline{\bm{A}}}
\newcommand{\meanBB}{\overline{\bm{B}}}
\newcommand{\meanJJ}{\overline{\bm{J}}}
\newcommand{\meanUU}{\overline{\bm{U}}}
\newcommand{\zzz}{\hat{\mbox{\boldmath $z$}} {}}

\newcommand{\meanBBhat}{\hat{\overline{\bm{B}}}}

\newcommand{\meanBhat}{\hat{\overline{B}}}

\newcommand{\AAA}{\bm{A}}
\newcommand{\aaa}{\bm{a}}
\newcommand{\BB}{\bm{B}}
\newcommand{\bb}{\bm{b}}
\newcommand{\JJ}{\bm{J}}
\newcommand{\jj}{\bm{j}}
\newcommand{\UU}{\bm{U}}
\newcommand{\vv}{\bm{v}}
\newcommand{\uu}{\bm{u}}
\newcommand{\oo}{\bm{\omega}}

\newcommand{\nab}{\mbox{\boldmath $\nabla$} {}}

\newcommand{\dd}{{\rm d} {}}
\newcommand{\const}{{\rm const}  {}}

\def\Rm{R_{\rm m}}
\def\Rey{\mbox{\rm Re}}

\def\St{\mbox{\rm St}}

\def\kf{k_{\rm f}}
\def\urms{u_{\rm rms}}
\def\Beq{B_{\rm eq}}
\def\etat{\eta_{\rm t}}

\newcommand{\alphaM}{\alpha_{\it M}}

\def\half{{\textstyle{1\over2}}}

\def\onethird{{\textstyle{1\over3}}}
\def\onesixth{{\textstyle{1\over6}}}

\newcommand{\yan}[5]{, ``#5,'' {\em Astron.\ Nachr.\ }{\bf #2}, #3-#4 (#1)}

\newcommand{\yana}[5]{, ``#5,'' {\em Astron.\ Astrophys.\ }{\bf #2}, #3-#4 (#1)}
\newcommand{\yanaS}[5]{, ``#5'' {\em Astron.\ Astrophys.\ }{\bf #2}, #3-#4 (#1)}

\newcommand{\ysph}[5]{, ``#5,'' {\em Solar Phys.\ }{\bf #2}, #3-#4 (#1)}

\newcommand{\ymn}[5]{, ``#5,'' {\em Monthly Notices Roy.\ Astron.\ Soc.\ }{\bf #2}, #3-#4 (#1)}

\newcommand{\ynat}[5]{, ``#5,'' {\em Nature }{\bf #2}, #3-#4 (#1)}

\newcommand{\sjfmS}[2]{, ``#2'' {\em J.\ Fluid Mech.} submitted (#1)}

\newcommand{\yjfm}[5]{, ``#5,'' {\em J.\ Fluid Mech.}{\bf #2}, #3-#4 (#1)}
\newcommand{\ypr}[5]{, ``#5,'' {\em Phys.\ Rev.\ }{\bf #2}, #3-#4 (#1)}
\newcommand{\ypre}[5]{, ``#5,'' {\em Phys.\ Rev.\ E }{\bf #2}, #3-#4 (#1)}
\newcommand{\ypreN}[4]{, ``#4,'' {\em Phys.\ Rev.\ }{\bf #2}, #3 (#1)}
\newcommand{\yprl}[5]{, ``#5,'' {\em Phys.\ Rev.\ Letters }{\bf #2}, #3-#4 (#1)}
\newcommand{\yprlN}[4]{, ``#4,'' {\em Phys.\ Rev.\ Letters }{\bf #2}, #3 (#1)}

\newcommand{\yprs}[5]{, ``#5,'' {\em Proc.\ Roy.\ Soc.\ Lond.\ }{\bf #2}, #3-#4 (#1)}
\newcommand{\yptrs}[5]{, ``#5,'' {\em Phil.\ Trans.\ Roy.\ Soc.\ }{\bf #2}, #3-#4 (#1)}

\newcommand{\yapj}[5]{, ``#5,'' {\em Astrophys.\ J.\ }{\bf #2}, #3-#4 (#1)}

\newcommand{\yapjl}[5]{, ``#5,'' {\em Astrophys.\ J.\ Letters }{\bf #2}, #3-#4 (#1)}
\newcommand{\ypp}[5]{, ``#5,'' {\em Phys.\ Plasmas }{\bf #2}, #3-#4 (#1)}

\newcommand{\ypf}[5]{, ``#5,'' {\em Phys.\ Fluids }{\bf #2}, #3-#4 (#1)}

\newcommand{\ygafd}[5]{, ``#5,'' {\em Geophys.\ Astrophys.\ Fluid Dynam. }{\bf #2}, #3-#4 (#1)}

\newcommand{\yjour}[6]{, ``#6,'' {\em #2} {\bf #3}, #4-#5 (#1)}
\newcommand{\yjourN}[5]{, ``#5,'' {\em #2} {\bf #3}, #4 (#1)}
\newcommand{\yjourNN}[4]{, ``#4,'' {\em #2}, p.~#3 (#1)}

\newcommand{\yproc}[7]{, ``#4,'' In {\em #5} (ed.\ #6), pp.\ #2-#3.\ #7 (#1)}

\newcommand{\sana}[2]{ ~#1~ ``#2,'' {\em Astron.\ Astrophys.} (submitted)}
\newcommand{\pana}[2]{ ~#1~ ``#2,'' {\em Astron.\ Astrophys.\ } (in press)}
\newcommand{\sapj}[2]{ ~#1~ ``#2,'' {\em Astrophys.\ J.} (submitted)}

\newcommand{\pmn}[2]{ ~#1~ ``#2,'' {\em Monthly Notices Roy.\ Astron.\ Soc.} (in press)}

\begin{document}

\title{Advances in theory and simulations of large-scale dynamos}


\author{Axel Brandenburg}


\institute{A. Brandenburg \at
Nordita, Roslagstullsbacken 23, 10691 Stockholm, Sweden\\
              Tel.: +46 8 5537 8707\\
              \email{brandenb@nordita.org}
}

\date{Received: date / Accepted: date}

\maketitle

\begin{abstract}
Recent analytical and computational advances in the theory of
large-scale dynamos are reviewed.
The importance of the magnetic helicity constraint is
apparent even without invoking mean-field theory.
The tau approximation yields expressions that show how the magnetic helicity
gets incorporated into mean-field theory.
The test-field method allows an accurate numerical determination of
turbulent transport coefficients in linear and nonlinear regimes.
Finally, some critical views on the solar dynamo are being offered
and targets for future research are highlighted.
\keywords{solar dynamo\and Sun\and magnetic fields\and magnetic activity}
\end{abstract}

\section{Introduction}
\label{intro}

Over the past 50 years significant progress has been made in understanding
the origin of the solar magnetic field.
In an important paper, Parker (1955) introduced the idea of mean magnetic
fields and identified the $\alpha$ effect as the crucial ingredient of
large-scale dynamos.
He also proposed and solved an explicit one-dimensional mean-field
model and found the migratory Parker dynamo wave.
This provided an important tool for understanding the effects of $\alpha$
and shear, and it led to useful estimates for the excitation conditions,
the cycle period, and the direction of field migration in solar and
stellar dynamo models.
However, Parker's work appeared at a time when it was still unclear whether
homogeneous fluid dynamos really exist.
These are dynamos of uniformly conducting matter, without insulating wires
that are thus susceptible to ``short circuits''.
In the years following Cowling's (1933) theorem, it remained doubtful
whether the Sun's magnetic field can be explained in terms of dynamo theory,
as originally anticipated by Larmor (1919).

In the paper on his famous theorem, Cowling (1933) concluded
``The theory proposed by Sir Joseph Larmor, that the magnetic field of
a sunspot is maintained by the currents it induces in moving matter,
is examined and shown to be faulty; the same result also applies for the
similar theory of the maintenance of the general field of Earth and Sun.''
Larmor (1934) responded that ``the self-exciting dynamo analogy is still,
so far as I know, the only foundation on which a gaseous body such as
the Sun could possess a magnetic field: so that if it is demolished
there could be no explanation of the Sun's magnetic field even remotely
in sight.''

Although the first qualitative ideas on homogeneous dynamos were proposed
nearly a hundred years ago, the resistance was immense; for historical
accounts see the reviews by Krause (1993) and Weiss (2005).
An important existence proof for homogeneous self-excited dynamos was
that of Herzenberg (1958), who showed, using asymptotic theory, that
dynamos work in a conducting medium where two rotors spin about axes that
lie in planes perpendicular to their direction of separation,
and inclined relative to each other by an angle between $90^\circ$
and $180^\circ$.
Such systems were later realized experimentally by
Lowes \& Wilkinson (1963, 1968).
In their experiments oscillations commonly occurred.
Those where thought to be some kind of nonlinear relaxation oscillations.
However, they used angles of less than $90^\circ$.
Indeed, when the relative angle between the rotors is between
$0^\circ$ and $90^\circ$, oscillatory solutions are expected
even from linear theory (Brandenburg et al.\ 1998).
Those solutions were not captured by the original analysis of
Herzenberg (1958), because he only looked for steady solutions.

The next important steps came with the development of mean-field
electrodynamics by Steenbeck, Krause, \& R\"adler (1966), who used
the first order smoothing approximation (or second order correlation
approximation) as a rigorous tool to compute $\alpha$ effect and turbulent
diffusivity in limiting cases.
Steenbeck \& Krause (1969) later produced global mean-field models
in spherical geometry and computed synthetic butterfly diagrams.
For an introduction to mean-field theory we refer to the
article by N.\ O.\ Weiss in this issue.

The technical tools made available by mean-field theory have stimulated
much of the research in the field during the 1970s.
However, during the 1980s a number of problems were discussed.
For example, doubts were raised whether turbulent magnetic diffusion
still works at large magnetic Reynolds numbers, $\Rm$; see work by
Knobloch (1978), Layzer et al.\ (1979) and Piddington (1981).
This problem applies equally to kinematic and nonlinear cases.
Regarding the kinematic $\alpha$ effect, Childress (1979) found that
in steady convection $\alpha$ decreases with increasing $\Rm$ like
$\Rm^{-1/2}$.
This result is now understood to be a common feature of steady flows
(R\"adler et al.\ 2002, R\"adler \& Brandenburg 2008), and is generally
not shared by unsteady (e.g.\ turbulent) flows (Sur et al.\ 2008).

The nonlinear problem was a focus of much of the work on dynamos during
the 1990s, and started with the work of Cattaneo \& Vainshtein (1991,
hereafter referred to as CV91)
who showed, using two-dimensional turbulence simulations, that
for $\meanBB^2\approx B_{\rm eq}^2$, $\etat$ decreases like $\Rm^{-1}$.
It was expected that a similar relation applies also to $\alpha$
(Vainshtein \& Cattaneo 1992, hereafter VC99), but this required
three-dimensional considerations.
Indeed, using uniform imposed fields,
Cattaneo \& Hughes (1996, hereafter CH96) showed that
$\alpha$ decays with increasing $\Rm$ like $\Rm^{-1}$.
These results were later understood to be due to the presence of
conservation laws for the mean squared vector potential, $\bra{\AAA^2}$,
in two dimensions and the magnetic helicity, $\bra{\AAA\cdot\BB}$,
in three dimensions Gruzinov \& Diamond (1994, hereafter GD94, 1995).
Here, $\AAA$ is the magnetic vector potential with $\BB=\nab\times\AAA$.
However, these conservation laws only tell us how much small-scale
magnetic field is being produced as the mean-field dynamo
produces large-scale field, such that $\bra{\AAA^2}$ (in two dimensions)
or $\bra{\AAA\cdot\BB}$ (in three dimensions) remain unchanged.
One still needs a theory that relates the corresponding small-scale
mean squared vector potential to the turbulent magnetic diffusion
in two dimensions or the small-scale magnetic helicity to the turbulent
diffusivity or the $\alpha$ effect in three dimensions.
This can be done using a corresponding mean-field equation
for these quantities.

The effect of such nonlinear dependencies of turbulent transport
coefficients on the dynamo can be quite dramatic.
In three dimensions, Gruzinov \& Diamond (1995)
showed that in the case of pseudo-vacuum boundary conditions
the saturation field strength of a dynamo with just helicity is
of the order of $\Rm^{-1/2}\Beq$.
This was also confirmed by simulations
(Brandenburg \& Dobler 2001, hereafter BD01;
Brandenburg \& Subramanian 2005a).
In the special case of periodic boundary conditions, however,
the field strength does not decline, but remains of the order of
$(\kf/k_1)^{1/2}\Beq$ (Brandenburg 2001, hereafter B01).
This is now well understood as being a consequence of magnetic helicity
evolution, which was soon applied to cases with shear
(Brandenburg et al.\ 2001, hereafter BBS01;
Blackman \& Brandenburg 2002, hereafter BB02)
in domains with periodic as well as open boundary conditions
(Brandenburg 2005, hereafter B05).
Magnetic helicity evolution has also been invoked to understand recent
simulations of convection by Tobias et al.\ (2008, hereafter; TCB08)
and K\"apyl\"a et al.\ (2008a, hereafter KKB08).

\Tab{Tsummary} summarizes a number of results that have been
obtained over the years.
These results may appear conflicting at first sight, but they are
in fact all explained by modern dynamo theory that takes magnetic
helicity evolution into account, and that allows for magnetic helicity
changes in the presence of losses through boundaries.
In the following we restrict ourselves to cases in Cartesian geometry,
but we note that important progress is now also being made in spherical
shell geometry where large-scale fields have been seen when rotation
is sufficiently rapid (Brown et al.\ 2007).

\begin{table}[t!]\caption{Summary of results obtained over the years.
The key to the references is given at the end of \Sec{intro}.
}\vspace{12pt}\centerline{\begin{tabular}{lll}
Result & Details & Reference \\
\hline
$\etat\sim\Rm^{-1}$ & 2-D periodic, $\meanB\sim\sin kx$ & CV91 \\
$\alpha\sim\Rm^{-1}$ & phenomenology, $\bra{\AAA\cdot\BB}$ conservation,
simulations with $\meanB=\const$ & VC92, GD94, CH96 \\
$\meanB^2/\Beq^2\sim\Rm^{-1}$ & helical turbulence, normal field b.c. & GD94, BD01\\
$\meanB^2/\Beq^2\sim\kf/k_1$ & helical turbulence, periodic domain & B01\\
$\meanB^2/\Beq^2\gg\kf/k_1$ & helical turb. with shear, periodic & BBS01, BB02\\
$\meanB^2/\Beq^2\sim0.5$ & helical turb. with horizontal shear,
normal field b.c. & B05\\
$\meanB^2/\Beq^2\ll0.5$ & convection with vertical shear,
normal field b.c. & TCB08\\
$\meanB^2/\Beq^2\sim0.5$ & convection with horizontal shear,
normal field b.c. & KKB08\\
\label{Tsummary}\end{tabular}}\end{table}

\section{Saturation phenomenology in a periodic box}
\label{SaturationPheno}

During the early phase of a strongly helical dynamo there can
be a phase during which the magnetic energy of the large-scale field
is still subdominant.
However, at later times the magnetic energy can redistribute itself from
small to large scales.
The fields that suffer minimal back-reaction from the Lorentz force
tend to be force-free at large scales.
Force-free fields are generally referred to as Beltrami fields.
Qualitatively speaking, the helical driving produces a helical
field at the driving scale, but because magnetic helicity cannot change,
helical field of opposite helicity must emerge at some other scale.
Simple arguments show that this can only happen at a larger scale
(Frisch et al.\ 1975; see also Brandenburg \& Subramanian 2005b).
To explain the evolution of the resulting large-scale magnetic field,
let us begin with the evolution equation of magnetic helicity,
\EQ
{\dd\over\dd t}\bra{\AAA\cdot\BB}=-2\eta\mu_0\bra{\JJ\cdot\BB},
\label{dABdt}
\EN
where angular brackets denote volume averages, $\eta$ is the
microscopic magnetic diffusivity, $\mu_0$ is the vacuum permeability,
and $\JJ=\nab\times\BB/\mu_0$ is the current density.
Next, we introduce horizontal averages denoted by overbars.
The direction over which we take these averages depends of course
on the direction in which the mean magnetic field chooses to align itself.
There are three equivalent possibilities, so let us assume that the field
shows a large-scale modulation in the $z$ direction.
In a periodic box the Beltrami field with the smallest wavenumber is then
of the form
\EQ
\meanBB=\meanBB(z,t)=\hat{B}(t)\left(\cos k_1z,\sin k_1 z, 0\right),
\label{coszsinzField}
\EN
where we have ignored the possibility of an arbitrary phase shift
in the $z$ direction.
Note that $\meanJJ(z,t)=-k_1\meanBB/\mu_0$ and $\meanAA(z,t)=-k_1^{-1}\meanBB$,
so the current and magnetic helicities have negative sign at large scales.
This is the situation when the small-scale driving has positive helicity.

Note that the definition of averaging automatically defines small-scale
(or fluctuating) magnetic fields as $\bb=\BB-\meanBB$, and likewise
for $\aaa=\AAA-\meanAA$ and $\jj=\JJ-\meanJJ$.
We can then split \Eq{dABdt} into contributions from large scales
and small scales, reorganize the equations in terms of $\bra{\meanBB^2}$
and $\bra{\bb^2}$, assume that, after the end of the kinematic phase
($t=t_{\rm s}$), $\bra{\bb^2}$ is approximately constant in time
(approximately equal to $\mu_0\bra{\rho\uu^2}$).
This yields (B01)
\EQ
k_1^{-1}{\dd\bra{\meanBB^2}\over\dd t}
=2\eta\kf\bra{\bb^2}-2\eta k_1\bra{\meanBB^2},
\EN
which has the solution
\EQ
\bra{\meanBB^2}=\bra{\bb^2}{\kf\over k_1}
\left[1-\mbox{e}^{-2\eta k_1^2(t-t_{\rm  s})}\right].
\label{meanBBpheno}
\EN
Thus, $\bra{\meanBB^2}$ saturates on a time scale $(2\eta k_1^2)^{-1}$,
i.e.\ the microscopic diffusion time based on the scale of the box.
This equation reproduces extremely well the saturation behavior in a
periodic box.
This equation also shows what happens if either the fluctuating field or the
mean field are not fully helical (Brandenburg et al.\ 2002).
For example, if the large-scale field is no longer fully helical, then
the ratio $\mu_0|\bra{\meanJJ\cdot\meanBB}|/\bra{\meanBB^2}$ will be less
then $k_1$, so we say that the {\it effective} value of $k_1$ will be smaller.
(Later on we refer to this value as $k_{\rm m}$.)
Thus, if the large-scale field is not fully helical,
but the small-scale field is still fully helical, then the effective
value of $k_1$ in the denominator of \Eq{meanBBpheno} decreases and
$\bra{\meanBB^2}$ can
be even somewhat higher than for periodic boundary conditions.
This is indeed the case for perfectly conducting boundary conditions,
which do not permit \Eq{coszsinzField} as a solution.
This is the reason why the effective value of $k_1$ is smaller, and
hence $\bra{\meanBB^2}$ is larger (Brandenburg \& Dobler 2002),
Conversely, if the small-scale field is not fully helical,
the effective value of $\kf$ is smaller, and so $\bra{\meanBB^2}$
is smaller (Maron \& Blackman 2002, Brandenburg et al.\ 2002).

We emphasize that in the considerations in this section we did
not invoke mean-field theory at all.
The slow-down during the final saturation stage is rather general and
it should be possible to describe this by a sufficiently detailed mean-field
theory.
This will be discussed briefly in the following section.

\section{Mean-field theory and transport coefficients}

In mean-field theory one considers the averaged induction equation.
The cross-product of the correlation of the fluctuations
$\uu=\UU-\meanUU$ and $\bb=\BB-\meanBB$, i.e.\ the mean electromotive
force, $\meanEMF=\overline{\uu\times\bb}$, provides an important term
in the averaged induction equation,
\EQ
{\partial\meanBB\over\partial t}=\nab\times\left(\meanUU\times\meanBB
+\meanEMF-\eta\mu_0\meanJJ\right).
\EN
A central goal of mean-field theory is to find expressions for
$\meanEMF$ in terms of mean-field quantities.
Quadratic correlations such as $\meanEMF$ are obtained using evolution equations
for the fluctuations, $\uu\equiv\UU-\meanUU$ and $\bb=\meanBB-\bb$.
A range of different approaches can be used to calculate the functional
form of the mean electromotive force, $\meanEMF=\overline{\uu\times\bb}$,
including the second order correlation approximation (SOCA), the
$\tau$ approximation, and the renormalization group procedure.
Common to both the SOCA and the $\tau$ approximation is the fact that
the linear terms in the evolution equations for the fluctuations are
solved exactly.
However, there is an important difference in that the $\tau$ approximation
starts by computing the time evolution of $\meanEMF$, so one begins with
\EQ
\partial\meanEMF/\partial t
=\overline{\dot{\uu}\times\bb}+\overline{\uu\times\dot{\bb}},
\label{dEMFdtAnsatz}
\EN
whereas under SOCA one uses primarily the induction equation by computing
$\meanEMF=\overline{\uu\times\bb}$, where $\uu$ is assumed given and $\bb$
is being solved using the Green's function for the induction equation.
In simple terms, this reduces to solving for
$\meanEMF=\overline{\uu\times\int\dot{\bb}\,\dd t}$.
This distinction is important because under the $\tau$ approximation the term
$\overline{\dot{\uu}\times\bb}$ leads immediately to a term of the form
$\overline{(\jj\times\meanBB)\times\bb}$ owing to the Lorentz force.
This expression leads to an important feedback by attenuating the
$\alpha$ effect by a term $\alphaM$, where, under the
assumption of isotropy, $\alphaM=\onethird\tau\overline{\jj\cdot\bb}$ is
the magnetic $\alpha$ effect.
Another important difference is that there is a natural occurrence of a
time derivative of $\meanEMF$.
Thus, compared with SOCA, which leads to
\EQ
\meanemf_i=\alpha_{ij}\meanB_j+\eta_{ijk}\meanB_{j,k}\;,
\label{EMF}
\EN
one now has
\EQ
\tau\partial\meanemf_i/\partial t+
\meanemf_i=\alpha_{ij}\meanB_j+\eta_{ijk}\meanB_{j,k}\;,
\label{dEMFdt}
\EN
where $\tau$ is a relaxation time, and a comma between indices denotes
a spatial derivative.
In \Eq{dEMFdt} the origin of the $\tau\partial\meanEMF/\partial t$ term
is clear in view of \Eq{dEMFdtAnsatz}, and it is instead the $\meanEMF$
term that is due to retaining nonlinear terms in the evolution equations
for $\uu$ and $\bb$.
In both cases these terms lead to the triple correlations that are then
approximated by $-\meanEMF/\tau$ on the right hand side.
After multiplying by $\tau$, this leads to the $\meanEMF$ term in
\Eq{dEMFdt}.

In the expressions above we have used the more general tensorial
forms of $\alpha$ effect and turbulent diffusion.
Scalar transport coefficients used before denote the isotropic
contributions of the $\alpha_{ij}$ and $\eta_{ijk}$ tensors, i.e.\
$\alpha=\onethird\delta_{ij}\alpha_{ij}$ and
$\eta_{\rm t}=\onesixth\epsilon_{ijk}\eta_{ijk}$.

Both SOCA and the $\tau$ approximation are rather primitive and their
merits has been discussed in some detail in the recent literature
(R\"adler \& Rheinhardt 2007, Sur et al.\ 2007).
The emergence of the $\overline{\jj\cdot\bb}$ term is qualitatively
a new feature that leads to a quantitative description of the saturation
of large-scale dynamos in periodic domains (Field \& Blackman 2002, BB02).
Furthermore, the emergence of an additional time derivative in \Eq{dEMFdt}
has been confirmed qualitatively using simulations (Brandenburg et al.\ 2004).
However, there is now also evidence for the occurrence of even higher
time derivatives in some cases (Hubbard \& Brandenburg 2008).

The time derivative in \Eq{dEMFdt} suppresses changes of mean-field
properties on timescales shorter than the turnover time $\tau$ of the
turbulence.
This is analogous to the occurrence of the Faraday displacement current
in the Maxwell equations, except that there the limiting velocity is
the speed of light, whereas here it is the rms velocity of the turbulence.
This changes the parabolic nature of the diffusion and dynamo equations
into hyperbolic wave equations (Blackman \& Field 2003,
Brandenburg et al.\ 2004).
This property is physically appealing, because it retains causality,
which means here that no mean-field pattern can propagate faster than
the rms velocity of the turbulence.

Similar to the suppression of fast temporal variations discussed above,
there is also a suppression of spatial variations on short length scales.
Indeed, \Eq{dEMFdt} takes the more accurate form
\EQ
\tau\partial\meanemf_i/\partial t+
\meanemf_i=\hat\alpha_{ij}\circ\meanB_j+\hat\eta_{ijk}\circ\meanB_{j,k},
\label{dEMFdtKernel}
\EN
where $\hat\alpha_{ij}$ and $\hat\eta_{ijk}$ are the components of
integral kernels and the circles denote a convolution.
Recent numerical work has now established that for driven turbulence
the integral kernels have an exponential form with a width given by
the inverse wavenumber of the energy-carrying eddies
(Brandenburg et al.\ 2008b).

This implies that mean-field theory should never produce rapid spatial
or temporal variations.
Conversely, the more complicated kernel formulation in \Eq{dEMFdtKernel}
can be avoided if the solutions are sufficiently smooth in space and time.
However, this is not always guaranteed, especially near boundaries.

Let us at this point also highlight the occurrence of another time
derivative in the mean-field equations.
Under the $\tau$ approximation, the
$\overline{(\jj\times\meanBB)\times\bb}$ term
leads to the emergence of a magnetic contribution to the $\alpha$ effect.
The full $\alpha$ effect is then written as
$\alpha=\alpha_{\rm K}+\alpha_{\rm M}$,
where $\alpha_{\rm K}$ is related to the kinetic helicity and
$\alpha_{\rm M}$ is related to the current helicity.
The latter obeys an evolution equation where the omission of the
time-derivative is often problematic, especially when $\Rm$ is large
and the mean divergence of current helicity fluxes vanishes.
Therefore, the more complete quenching formula with extra effects
included takes the form (see, e.g., Brandenburg 2008),
\EQ
\alpha={\alpha_0+\Rm\left(
\eta_{\rm t}{\mu_0\meanJJ\cdot\meanBB\over B_{\rm eq}^2}
-{\nab\cdot\meanFF_{\rm C}\over2 k_{\rm f}^2B_{\rm eq}^2}
-{\partial\alpha/\partial t\over2\eta_{\rm t} k_{\rm f}^2}\right)
\over1+\Rm\meanBB^2/B_{\rm eq}^2}.
\label{QuenchExtra2}
\EN
Although this equation can be written as an evolution equation, in practice
there is a computational advantage in solving the time-derivative term
implicitly; see Brandenburg \& K\"apyl\"a (2007).
The properties of such a ``dynamical'' $\alpha$ quenching formula have been
studied in a number of recent papers including Kleeorin et al.\ (2000),
Field \& Blackman (2002), BB02, and Brandenburg \& Subramanian (2005a).

\section{The test-field method}

In the last few years a new and reliable method for calculating the
$\alpha_{ij}$ and $\eta_{ijk}$ tensor coefficients has become available.
This method is known as the test-field method and was developed by
Schrinner et al.\ (2005, 2007) to calculate all tensor components
from snapshots of simulations of the geodynamo in a spherical shell.
This method was later applied to time-dependent turbulence in
triply-periodic Cartesian domains, both with shear and no helicity
(Brandenburg 2005; Brandenburg et al.\ 2008a) as well as without shear,
but with helicity (Sur et al.\ 2008; Brandenburg et al.\ 2008b), and
also with both (Mitra et al.\ 2008a).

\subsection{The essence of the test-field method}

In the test-field method one solves an additional set of three-dimensional
partial differential equations for vector fields $\bb^{pq}$, where the labels
$p=1,2$ and $q=1,2$ correspond to different pre-determined one-dimensional
test fields $\meanBB^{pq}$.
The evolution equations for $\bb^{pq}$ are derived by subtracting the
mean-field evolution equation from the evolution equation for $\BB$.
These equations are {\it distinct} from the original induction equation
in that the curl of the resulting mean electromotive force is subtracted.

The test-field method has recently been criticized by Cattaneo \& Hughes (2008)
on the grounds that the test fields are arbitrary pre-determined mean fields.
They argue that the resulting turbulent transport coefficients
will only be approximations to the true values unless the test fields
are close to the actual mean fields.
Mitra et al.\ (2008a) have reviewed arguments supporting the
validity of the test-field method:
(i) the test-field method correctly reproduces a vanishing growth
rate in saturated nonlinear cases (Brandenburg et al.\ 2008c);
(ii) in the time-dependent case, the test-field method
correctly reproduces also a non-vanishing growth rate.
In that case one must write Eq.~(\ref{EMF}) as a convolution in time
(Hubbard \& Brandenburg 2008);
(iii) for the Roberts flow with a mean field of Beltrami type,
the $\alpha_{ij}$ tensor is anisotropic and has an additional component
proportional to $\meanB_i\meanB_j$ that tends
to quench the components of the isotropic part of $\alpha_{ij}$.
The same $\alpha_{ij}$ tensor also governs the evolution of a mean
passive vector field.
It turns out that the fastest growing passive vector field is then
phase-shifted by 90 degrees relative to the one that caused the
quenching and thus the quenched form of $\alpha_{ij}$.
This result has been confirmed both numerically and
using weakly nonlinear theory (Tilgner \& Brandenburg 2008).
We discuss this case further in section \ref{Quenching}.

\subsection{$\Rm$-dependence of the kinematic values of $\alpha$ and $\etat$}

\begin{figure}\begin{center}
\includegraphics[width=\columnwidth]{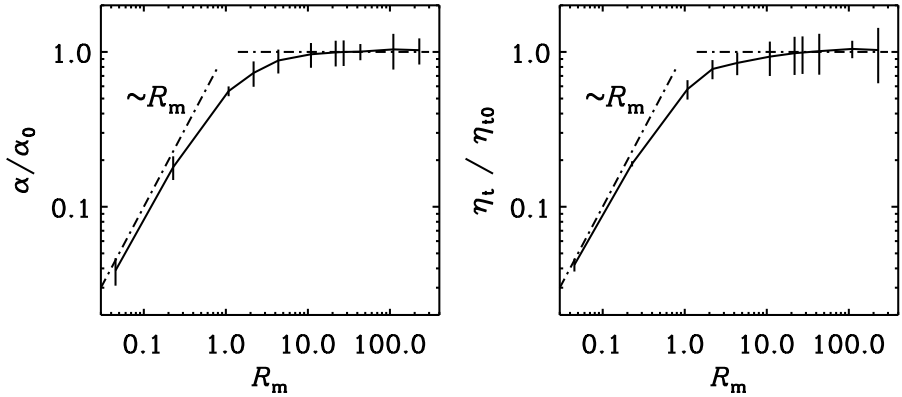}
\end{center}\caption[]{
Dependence of the normalized values of $\alpha$ and $\eta_{\rm t}$
on $R_{\rm m}$ for $\mbox{Re}=2.2$.
The vertical bars denote twice the error estimated by averaging over
subsections of the full time series.
The run with $\Rm=220$ ($\Rey=2.2$) was done at a resolution of
$512^3$ meshpoints.
Adapted from Sur et al.\ (2008).
}\label{Re2}\end{figure}

Using the test-field method it has, for the first time, become possible to
obtain reliable estimates not only for the $\alpha$ effect, but in particular
also for the turbulent magnetic diffusivity.
Restricting ourselves to the case of horizontal ($xy$) averages,
the mean fields depend only on  $z$ and $t$.
All components of $\meanB_{j,k}$ can therefore be expressed in terms
of those of $\meanJJ(z,t)$, and the relevant components of $\eta_{ijk}$
reduce to a rank-2 tensor, $\eta_{ij}$.
In that case, $\etat=\half(\eta_{11}+\eta_{22})$.
We present the $\Rm$ dependences of $\alpha$ and $\etat$ in normalized forms
using the SOCA results for homogeneous isotropic turbulence as reference values,
\EQ
\alpha_0=-\onethird\tau\overline{\oo\cdot\uu},\quad
\eta_{\rm t0}=\onethird\tau\overline{\uu^2}
\quad\mbox{(SOCA, linear)}.
\label{SOCAlinear}
\EN
It turns out that, in the kinematic regime, $\alpha_0$ and $\eta_{\rm t0}$
are remarkably close to the numerically determined values of $\alpha$
and $\eta_{\rm t}$ in the range $1<\Rm<200$ considered in the study of
Sur et al.\ (2008); see \Fig{Re2}.
For $\Rm<1$, both $\alpha$ and $\eta_{\rm t}$ increase linearly
with $\Rm$.
In the cases considered here we have assumed that the turbulence is
fully helical, so $\overline{\oo\cdot\uu}\approx\kf\overline{\uu^2}$,
and that the Strouhal number, $\St\equiv\tau\urms\kf$ is approximately
equal to unity (Brandenburg \& Subramanian 2005c, 2007).
Of course, for $\Rm<1$ this is not the case and then
$\tau\approx(\eta\kf^2)^{-1}$ is a better estimate.
This explains the linear increase of $\alpha$ and $\etat$ for $\Rm<1$.

\subsection{Scale-dependence of $\alpha$ and $\etat$}

Using the test-field method, it has now also been possible to
determine what happens if there is poor scale separation, for example
if the scale of the mean field is only 2--5 times bigger than the
scale of the energy-carrying eddies.
In that case one can not longer write the electromotive force in terms
of products of $\alpha$ and the mean field or $\etat$ and the mean current
density, but one has to write them as convolutions with corresponding
integral kernels (e.g.\ Brandenburg \& Sokoloff 2002).
In Fourier space, a convolution corresponds to a multiplication.
In the test-field method we use only harmonic test fields with a
single wavenumber, so we can use this method to calculate
$\alpha$ and $\eta_{\rm t}$ separately for each wavenumber and obtain
the integral kernels via Fourier transformation.

Not surprisingly, it turns out that $\alpha$ and $\eta_{\rm t}$
decrease with decreasing scale, i.e.\ with increasing values of
$k/\kf$, where $k$ is the wavenumber of a particular Fourier mode
of the field.
In fact, by calculating $\alpha$ and $\eta_{\rm t}$ for test-fields
of different wavenumber $k$, one finds that for isotropic turbulence,
$\alpha$ and $\eta_{\rm t}$ have Lorentzian profiles of the form
\EQ
\alpha(k)={\alpha_0\over1+(a_\alpha k/\kf)^2},\quad
\eta_{\rm t}(k)={\eta_{\rm t0}\over1+(a_\eta k/\kf)^2},
\EN
where $a_\alpha$ and $a_\eta$ are factors of order unity;
Brandenburg et al.\ (2008b) find $a_\alpha\approx1$ and $a_\eta\approx0.5$.
However, for shear-flow turbulence Mitra et al.\ (2008a) find
$a_\alpha\approx a_\eta\approx0.7$.

In periodic domains the Fourier transforms of $\alpha(k)$ and
$\eta_{\rm t}(k)$ correspond to the integral kernels introduced in
\Eq{dEMFdtKernel}.
They are of exponential form, i.e.,
\EQ
\hat\alpha(z-z')=\half a_\alpha\alpha_0\kf\sim\exp(-\kf|z-z'|/a_\alpha)
\EN
and
\EQ
\hat\etat(z-z')=\half a_\eta\eta_{\rm t0}\kf\sim\exp(-\kf|z-z'|/a_\eta).
\EN

It is important to realize that the test-field method is a tool
to analyze the velocity field that is giving rise to $\alpha$ and
$\eta_{\rm t}$ effects.
By applying the test-field method to the case where the induction
equation is solved together with the momentum and continuity
equations, one can analyze the nonlinear case for one specific
value of $\meanBB$.
We emphasize that the test field does not enter the momentum
equation in any way.
This will be discussed next.

\subsection{Quenching for equipartition-strength fields}
\label{Quenching}

Once the magnetic field has become sufficiently strong, $\alpha$ and
$\eta_{\rm t}$ will become anisotropic, even though the turbulence was
originally isotropic.
If the anisotropy is only due to $\meanBB$,
the tensors $\alpha_{ij}$ and $\eta_{ijk}$ are of the form
\EQ
\alpha_{ij}(\meanBB)=\alpha_1(\meanBB)\delta_{ij}
+\alpha_2(\meanBB)\meanBhat_i\meanBhat_j, \label{alpten}\EN
\label{alpten}
\EQ
\eta_{ij}(\meanBB)=\eta_1(\meanBB)\delta_{ij}
+\eta_2(\meanBB)\meanBhat_i\meanBhat_j,
\label{etaten}
\EN
where $\meanBBhat=\meanBB/|\meanBB|$ is the unit vector of the mean field.

\begin{figure}\begin{center}
\includegraphics[width=\columnwidth]{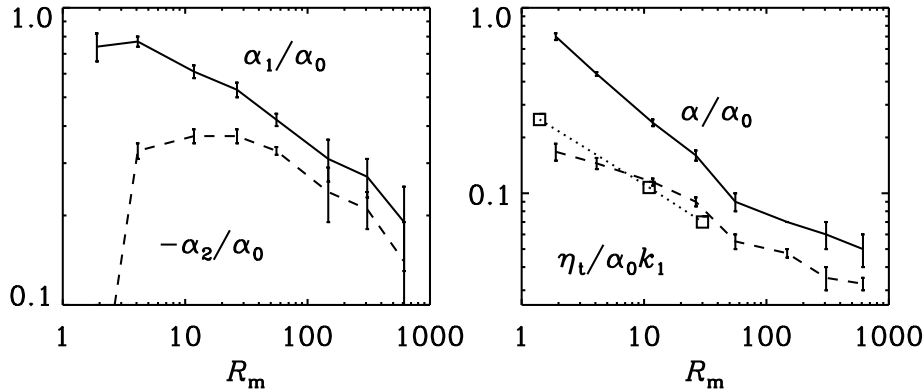}
\end{center}\caption[]{
$\Rm$ dependence of $\alpha_1$ and $-\alpha_2$ ({\it left}) and of
$\alpha$ and $\eta_{\rm t}k_1$ ({\it right})
for equipartition-strength fields, $|\meanBB|=O(B_{\rm eq})$.
The mutual approach of  $\alpha_1$ and $-\alpha_2$ illustrates how
$\alpha$ quenching is accomplished, and the mutual approach of
$\alpha$ and $\eta_{\rm t}k_1$ illustrates by how much the quenching
has to proceed.
}\label{palpeta}\end{figure}

For equipartition-strength fields, $|\meanBB|=O(B_{\rm eq})$,
the $\Rm$ dependence of $\alpha_1$, $\alpha_2$, $\eta_1$, and
$\eta_2$ has been determined by Brandenburg et al.\ (2008c).
It turns out that $\alpha_1$ and $\alpha_2$ have opposite signs
(\Fig{palpeta}), so when $\alpha_{ij}$ is applied to the actual
mean field we have
\EQ
\alpha_{ij}B_j=(\alpha_1+\alpha_2)B_i.
\EN
This shows that the $\alpha$ effect is magnetically quenched by the
suppressing effect of $\alpha_2$ on $\alpha_1$ due to its opposite sign.
However, even though the value of $\alpha_1+\alpha_2$ decreases with
increasing values of $\Rm$, it is only quenched down to values comparable
to the value of $\eta_1k_1$ if $|\meanBB|=O(B_{\rm eq})$; see \Fig{palpeta}.
This becomes obvious by looking at the expression for the linear growth
rate,
\EQ
\lambda=(\alpha_1+\alpha_2)k_{\rm m}-(\eta+\eta_1+\eta_2)k_{\rm m}^2,
\EN
where $k_{\rm m}=\mu_0\bra{\meanJJ\cdot\meanBB}/\bra{\meanBB^2}$
is the effective wavenumber of the mean field.
We note however that the use of $\lambda$ is only permissible because
$\meanBB^2$ and $\meanJJ\cdot\meanBB$ are spatially uniform for
$\alpha^2$ dynamos in a periodic domain.
For a forcing function with positive helicity we have $\kf>0$,
and so $k_{\rm m}<0$.
Moreover, for fully helical mean fields we have $k_{\rm m}=-k_1$.
In the saturated state, the growth rate must be zero,
which means then that $\alpha_1+\alpha_2$ must become comparable to
$(\eta+\eta_1+\eta_2)k_{\rm m}$.

The occurrence of the $\meanBhat_i\meanBhat_j$ term in \Eq{alpten} and the
negative sign of $\alpha_2/\alpha_1$ have been confirmed independently
by observing that the velocity field of a saturated dynamo can itself
lead to dynamo action for a passive vector field obeying a kinematic
induction equation.
Such an observation was first made by Cattaneo \& Tobias (2008) in a
convection-driven small-scale dynamo and later by
Tilgner \& Brandenburg (2008) for the Roberts (1972) flow dynamo, where
$\uu=\kf\psi\zzz+\nab\times\psi\zzz$ with
$\psi=(u_0/k_1)\cos k_1x\cos k_1y$ and $k_{\rm f}=\sqrt{2}k_1$.
As in the case of helical isotropic turbulence in a triply-periodic domain
the solutions for $\meanBB$ are also here Beltrami fields of the form
$\meanBB=(\cos k_1z,\sin k_1z, 0)$, where $k_0$ is the
horizontal wavenumber of the helices of the Roberts flow.

The resulting matrix $\meanBhat_i\meanBhat_j$ has eigenvalues 1 and 0.
In the saturated state, the eigenfunction corresponding to eigenvalue 0
is $\tilde{\meanBB}=(\sin k_1z,-\cos k_1z, 0)$ and has the growth rate
$\lambda=\alpha_1(\meanBB)k_1-[\eta_1(\meanBB)+\eta]k_1^2$, which is
positive, even after $\meanBB$ has reached saturation.
This corresponds to continued exponential growth of $\tilde{\meanBB}$,
which confirms the original finding based on the test-field method.

The results obtained using the test-field methods should of course be of
predictive value to be useful.
The application to a passive vector field discussed above is one example
where the result for the full nonlinear $\alpha$ tensor was used to predict
the evolution of the passive vector field.
Another example is the case of rigidly rotating convection.
Using the test-field method, K\"apyl\"a et al.\ (2008b) noticed that with
increasing rotation rate $\alpha$ increases and $\etat$ decreases.
This led to the prediction that there should be $\alpha^2$ dynamo action
(i.e.\ without any shear!) for sufficiently rapid rotation.
This was later confirmed using direct simulations (K\"apyl\"a et al.\ 2008c).

\section{Three paradigm shifts revisited}

Let us now turn attention to the Sun.
Solar dynamo theory has experienced arguably three major paradigm shifts
since its broad initial acceptance during the 1970s.
Inevitably, these paradigm shifts have brought the modelling further away
from the original ideas that were based on dynamo theory.
At the same time solar dynamo theory has lost much of its
initial rigor that dynamo theory used to be based on, i.e.\ the
profiles of $\alpha$ and $\etat$ are no longer calculated, but
are considered freely adjustable.
The same is true of the magnetic quenching properties of these profiles.
It its therefore important that the motivation for such departures
from  the original theory are well justified.
In the following we discuss and comment on each of the three paradigm shifts.

\subsection{Magnetic buoyancy: from distributed dynamos to the overshoot layer}

In an influential paper by Spiegel \& Weiss (1980), a number of different
aspects led to the suggestion that the solar dynamo operates at the
base of the convection zone.
One of the arguments concerned the rapid rise of magnetic flux tubes
from the bulk of the convection zone.
Subsequent simulations, however, have demonstrated a strongly opposing
effect due to turbulent magnetic pumping (Brandenburg \& Tuominen 1991;
Nordlund et al.\ 1992; Brandenburg et al.\ 1996; Tobias et al.\ 1998).
It appears, therefore, that magnetic buoyancy might not constitute
a problem for the dynamo, even though its effects are clearly visible in
regions where the field is strong.
An example is Fig.~10 of Brandenburg et al.\ (1996), where the strongest
tube is just ``hovering'' at the same height in a balance between magnetic
buoyancy and downward pumping.

\subsection{Helioseismology: overshoot layer and flux-transport dynamos}

The idea of dynamos operating in the overshoot layer was soon reinforced
when it became evident that in the bulk of the convection zone the
{\it radial} differential rotation, which is important for the mean-field
dynamo, is small.
At the time, the strongest shear was believed to occur at the bottom of
the convection zone.
The positive value of the radial differential rotation in this layer,
which is now called the tachocline, together with an $\alpha$ effect of
opposite sign relative to what it is in the bulk of the convection zone,
could explain the equatorward migration of the sunspot belts
(DeLuca \& Gilman 1986, 1988, R\"udiger \& Brandenburg 1995).
However, as with all models that have a positive radial angular velocity
gradient, also these models have the wrong phase relation, i.e.\
the radial and toroidal mean fields are in phase and not in antiphase,
as observed (Yoshimura 1976, Stix 1976).
However, the phase relation may not pose a serious problem
(Sch\"ussler 2005).

Another possibility is that the dynamo could operate with spatially
disjoint induction layers: an $\alpha$ effect with the usual sign near
the surface, and positive radial shear at the bottom of the
convection zone, coupled by meridional circulation.
This led to the now popular idea of flux-transport dynamos where the
meridional circulation is chiefly responsible for the equatorward
migration of the toroidal flux belts (see article by M.\ Dikpati in this
issue).
However, in recent years it became clear that in the outer 5\% of the Sun
by radius there is strong negative radial shear (Benevolenskaya et al.\
1999), which could in principle also explain the equatorward migration
in the framework of conventional solar dynamo theory (B05).
On the other hand, such a theory also faces problems of its own,
for example the latitudinal width of the flux belts is expected to
be only a few times bigger than the depths of the supergranulation
layer (Brandenburg \& K\"apyl\"a 2007), which would be too small.

\subsection{Catastrophic quenching: interface and flux-transport dynamos}

The possibility of catastrophic quenching led Parker (1993) to
propose the so-called interface dynamo where the magnetic field
would be weak in the bulk of the convection zone, so as to avoid
catastrophic quenching.
However, as discussed in the present paper, catastrophic quenching is
{\it always} a serious possibility, even for interface dynamo,
which means that magnetic helicity fluxes are needed to alleviate it.
One might well imagine that it is easier to shed magnetic helicity
when the dynamo operates closer to the surface.
However, such models have not yet been investigated in sufficient detail.

In conclusion, there are now reasons to believe that all three
paradigm shifts are problematic and may need to be reconsidered.
An alternative proposal would be that the solar dynamo operates
in the bulk of the convection zone, just as anticipated originally
in the 1970s, and that the near-surface shear layer may play an
important role in shaping the solar dynamo wave (B05).

\section{Implications and open problems}

In future work it will be important to improve
our understanding of the solar dynamo, in particular its location within
the Sun, its 22 year period, and the origin of the equatorward migration
of the sunspot belts.
As discussed in the previous section,
current thinking places the solar dynamo in the tachocline, i.e.\ the bottom
of the convection zone where the internal angular velocity turns from
nearly uniform in the interior to non-uniform in the convection zone.
The idea is that the field strength there exceeds the equipartition
value by a factor of 100 (D`Silva \& Choudhuri 1993), but such a
strong field has not yet been reported based on turbulent
three-dimensional dynamo simulations.
Observationally not much can be said yet, because such fields would be
below current helioseismological detection limits.
On the theoretical side, a serious problem is that one
assumes a turbulent magnetic Prandtl number of 100, instead of 1,
which is predicted by theory and simulations (Yousef et al.\ 2003).
Such considerations neglect however the turbulent viscosity
associated with the Maxwell stress of small-scale magnetic fields.
Clearly, any {\it ad hoc} modifications of the theory are the result of
trying to make the models reproduce the observations.
However, at the same time such models ignore some important findings
regarding the nonlinear behavior of the mean-field dynamo effect
at large magnetic Reynolds numbers.
Recent research has provided new detailed insights that
should be followed up using more realistic settings such as
spherical shell geometry.

There are several mechanisms proposed for explaining the cause of the
equatorward migration of magnetic activity belts at low solar latitudes.
Is it the rather feeble meridional circulation, as assumed in the now
popular flux transport models (Dikpati \& Charbonneau 1999),
even though one has to assume unrealistic values of the {\it turbulent}
magnetic Prandtl number, or is it perhaps the near-surface shear layer,
which would have indeed the right sign, as emphasized in B05.
To clarify things, future research may proceed along two parallel
strands; one is connected with the development and exploitation of
models in spherical geometry, and the other one is connected with
unresolved problems that can be addressed in Cartesian configurations.
In the following we list detailed steps of future research.

{\it Catastrophic quenching in spherical shells}.~
Catastrophic quenching behavior has still not yet been demonstrated
convincingly in closed spheres or spherical shell sectors using, e.g.,
perfectly conducting boundary conditions and forced turbulence.
Some work in this direction has already been done
(Brandenburg et al.\ 2007, Mitra et al.\ 2008b),
but the resolution is limited and the results not yet entirely conclusive.

{\it Testfield method in spherical geometry}.~
The test-field method needs to be re-examined in spherical coordinates.
Originally the test-field method was developed in connection
with full spheres, and then the test fields consisted of field components
of constant value or constant slope.
However, only afterwards it became clear that the scale (or wavenumber)
of the field components must be the same for one set of all tensor
components, and so it is necessary to work with spherical harmonic
functions as test fields.
In other words, constant and linearly varying field components are problematic.

{\it Dynamo in open shells with and without shear}.~
To verify our understanding of the saturation process of large-scale
dynamos it is important to calculate, at different magnetic Reynolds numbers,
the late stages of magnetic field evolution with open boundary conditions in
spherical shells or shell sectors with and without shear.
One expects low saturation amplitudes with energies of the mean magnetic
field being inversely proportional to the magnetic Reynolds number
in the absence of shear, but of order unity in the presence of shear.
The shear is here critical, because shear is responsible for the local
driving of small-scale magnetic helicity fluxes (Vishniac \& Cho 2001;
Subramanian \& Brandenburg 2004, 2006).

{\it Alpha effect from convection}.~
The calculation of the $\alpha$ effect in convective turbulence is
at the moment unclear.
For unstratified convection with an imposed field Cattaneo \& Hughes (2006)
find that $\alpha$ diminishes for large magnetic Reynolds numbers,
even for kinematically weak magnetic fields.
With stratification, on the other hand, K\"apyl\"a et al.\ (2008b) find
values of $\alpha$ that are compatible with those from simple estimates.
They used the test-field method while Cattaneo \& Hughes (2006) used an
imposed field and estimate $\alpha$ as the ratio between the resulting
field-aligned electromotive force and the imposed field.
However, at large magnetic Reynolds number there is dynamo action
producing also a mean field that might exceed the imposed field and
thereby modify the estimate for $\alpha$.
Another possible reason for the discrepancy could be related to the
presence or absence of stratification, because $\alpha$ is expected to
be proportional to the local gradient of density and turbulent velocity
(Steenbeck et al.\ 1966).
In unstratified Boussinesq convection the density is constant and the
turbulent velocity only changes near boundaries.
However, boundary effects could contribute to driving an $\alpha$ effect
(Giesecke et al.\ 2005).
Another problem could be poor scale separation, in which case the
electromotive force is not just proportional to $\alpha$ and it becomes
mandatory to use the integral kernel formulation instead
(Brandenburg et al.\ 2008b).

{\it Convective dynamos in spherical shells} are now widely studied
(Brun et al.\ 2004; Browning et al.\ 2006; Brown et al.\ 2007).
It would be useful to compare the resulting magnetic fields with
corresponding forced turbulence simulations in spherical shells and
see whether contact can be made with improved mean-field models.
This may require careful considerations of the scale-dependence of the
turbulent transport coefficients.

{\it Dynamos driven by magnetic instabilities}.~
There is now quite a number of studies looking at possibilities where the
flows driving the dynamo are due to the resulting magnetic field itself,
and are driven by magnetic instabilities.
Examples include magnetic buoyancy instabilities and the
magneto-rotational instability.
For example, the turbulence in accretion discs is believed to be driven
by the magnetorotational instability.
This was one of the first examples showing cyclic dynamo action somewhat
reminiscent of the solar dynamo (Brandenburg et al.\ 1995), and it
was believed to be a prototype of magnetically driven dynamos
(Brandenburg \& Schmitt 1998; R\"udiger \& Pipin 2000;
R\"udiger et al.\ 2001; Blackman \& Field 2004).
In the mean time, another example of a magnetically driven dynamo has
emerged, where magnetic buoyancy works in the presence of shear and
stratification alone (Brummell et al.\ 2002; Cline et al.\ 2003a,b;
Cattaneo et al.\ 2006).
This phenomenon may be superficially similar to a magnetically dominated
version of the shear--current effect (Rogachevskii \& Kleeorin 2003, 2004).
With the test-field method one is now in a good position to identify the
governing mechanism by determining all components of the $\alpha$ and
$\etat$ tensors.

{\it Magnetic flux concentrations near the surface}.~
In the conventional picture, active regions and sunspots are thought to
emerge as a result of magnetic flux tubes breaking through the surface.
Given that it is difficult to imagine such tubes rising unharmed all the
way from the bottom of the convection zone over so many pressure scale
heights, one must test alternative
scenarios in which the emergence of active regions and sunspots
can be explained as the result of flux concentrations from local
dynamo action via negative turbulent magnetic pressure effects
(Kleeorin \& Rogachevskii 1994) or turbulent flux collapse
(Kitchatinov \& Mazur 2000).
Clearly, the underlying effects need to be established numerically
and corresponding mean-field models need to be solved to make direct
contact with simulations.

{\it CME-like features above the surface}.~
Given that virtually all successful large-scale dynamos at large magnetic
Reynolds numbers are now believed to shed small-scale magnetic helicity,
it is important to analyze the nature of the expelled magnetic field in
simulations that couple to a simplified version of the lower solar wind.
It is possible that the magnetic field above the surface and
in the lower part of the solar wind might
resemble coronal mass ejections (CMEs), in which case more detailed
comparisons with actual coronal mass ejections would be beneficial.

{\it Solar cycle forecast}.~
Among the popular applications of solar dynamo theory and solar
magnetohydrodynamics
are solar cycle predictions, solar subsurface weather, and space weather.
Also of interest are predictions of solar activity during its first
500 thousand years.
This has great relevance for predicting the loss of volatile elements
from the Earth's atmosphere, for example, and for understanding the
conditions on Earth during the time when life began colonizing the planet.
In this connection it is important to calculate the deflection
of cosmic ray particles by the Sun's magnetic field and on
the scale of the galaxy which is relevant for galactic cosmic rays
(Svensmark 2007a,b).
However, such studies would not be very meaningful unless some of the
earlier projects in this list have resulted in a solar dynamo model
that is trustworthy from a theoretical and a practical viewpoint.

{\it Applications to laboratory liquid sodium dynamos}.~
Unexpected beneficial insights have come from recent laboratory dynamo
experiments.
Unlike numerical dynamos, experimental liquid metal dynamos are able to
address the regime of rather low values of the magnetic Prandtl number
of the order of $10^{-5}$, which is interesting in connection with solar
and stellar conditions.
At the same time the magnetic Reynolds number can be
large enough (above 100) to allow for dynamo action.
The Cadarache experiment is particularly interesting.
Simulations of such a flow have been attempted by various groups using the
Taylor-Green flow as a model (Ponty et al.\ 2004, 2005; Mininni et al.\ 2005;
Brandenburg \& K\"apyl\"a 2007).
Again, the nature of the resulting dynamo effect has not yet been
elucidated.
It would be useful to analyze the resulting flows using the test-field
method.
One may hope that such work can teach us important lessons about
large-scale and small-scale dynamos at low magnetic Prandtl number
(Schekochihin et al.\ 2005; Iskakov et al.\ 2007), which is
relevant to the Sun, but hard to address numerically with the
currently available computing capabilities.
Another relevant application is precession-driven dynamos (Tilgner 1999),
where it might be useful to consider this process for a range of different
geometries.

\section{Conclusions}

Looking back at some of the problems that dynamo theory was facing during
the early years, we can say that a good deal of them have now been solved.
For example the issue of turbulent magnetic diffusivity at large magnetic
Reynolds numbers has now been addressed rather convincingly for values
of $\Rm$ up to 200.
Such a result has only recently become possible with the development of
the test-field method.
At this point we have no evidence that this result may change for larger
values of $\Rm$.
Similar statements can be made about $\alpha$, where it is now reasonably
clear that in the {\it kinematic} regime $\alpha$ approaches a constant
value for $1\leq\Rm\leq200$.
It should be emphasized that these results hold for forced turbulence and
one must expect them to be different in cases of naturally forced turbulence
such as convection or flows driven by magnetic instabilities such as
the magneto-rotational instability (Brandenburg 2008) or the magnetic
buoyancy instability (Brandenburg \& Schmitt 1998; Thelen 2000).

Much larger values of $\Rm$ of $2\times10^5$ have been obtained for
the special case of the Galloway-Proctor flow for which $\alpha$
shows irregular sign changes with $\Rm$ (Courvoisier et al.\ 2006).
This flow is a time-dependent version of the Roberts flow where the
pattern wobbles in the plane with given amplitude and frequency.
Expressions of the form \eq{SOCAlinear} do not apply in this case where
the correlation time is infinite (R\"adler \& Brandenburg 2008).
In that sense the Galloway-Proctor flow is quite different from
a turbulent flow.
Asymptotic behavior for large $\Rm$ is only possible for sufficiently
large amplitude and/or frequency of the wobbling motion.

In the nonlinear case equally dramatic progress has been made in just
the past few years.
While it has long been clear that in closed domains $\alpha$ will be
quenched down to values that depend on the quenched value of $\etat$
and on the effective wavenumber of the mean field,
it remained unclear what the quenched value of $\etat$ is.
Recent evidence points to a suppression by a factor of 5 when $\Rm$
is increased from 2 to 600 (Brandenburg et al.\ 2008c).
However, this value may depend on circumstances and could be slightly
less strong in the presence of shear (K\"apyl\"a \& Brandenburg 2008).

In open domains there is the possibility that the resulting magnetic
field strength can still decrease to catastrophically quenched values
unless there is a finite divergence of the magnetic helicity flux
(Brandenburg \& Subramanian 2005a).
Such a flux can be driven efficiently in the presence of shear.
In order for this mechanism to operate, the contours of constant
shear velocity must cross the boundaries (KKB08, Hughes \& Proctor 2008),
which explains the lack of large-scale fields in simulations with horizontal
shear and periodic boundary conditions in that direction (Tobias et al.\ 2008).

There is clearly a long way to go before the solar dynamo problem
can be addressed in full.
There is hardly any doubt that the inclusion of magnetic helicity fluxes
will be important, but the precise functional form of the magnetic
helicity flux needs to be confirmed numerically.
In particular, the possible dependencies of the fluxes on $\meanBB$
and $\Rm$ are not well understood at present.

\begin{acknowledgements}
I would like to acknowledge Sasha Kosovichev and the other members of
the team on Subphotospheric Dynamics of the Sun at the International
Space Science Institute in Bern for providing an inspiring atmosphere.
This work was supported in part by the Swedish Research Council.
\end{acknowledgements}


\end{document}